\documentclass[journal]{IEEEtran}
\usepackage{amsmath,graphicx}

\usepackage{setspace,tabularx}
\usepackage[tight,footnotesize]{subfigure}
\usepackage{times}
\usepackage{adjustbox}
\usepackage{algorithm}
\usepackage{amsfonts}
\usepackage{amsmath}
\usepackage{amssymb}
\usepackage{array}
\usepackage{boxedminipage}
\usepackage[center]{caption}
\usepackage{cite}
\usepackage{color}
\usepackage{cuted}
\usepackage{floatflt}
\usepackage{float}
\usepackage{graphics}
\usepackage{graphicx}
\usepackage{indentfirst}
\usepackage{latexsym}
\usepackage{hyperref}
\hypersetup{colorlinks=true,linkcolor=black,filecolor=black,urlcolor=black,citecolor=black}
\usepackage{multirow}
\usepackage{picinpar}
\usepackage{placeins}
\usepackage{psfrag}
\usepackage{soul}
\usepackage{subfigure}
\usepackage[usestackEOL]{stackengine}
\usepackage{theorem}
\usepackage[flushleft]{threeparttable}
\usepackage{wrapfig}
\usepackage{mathtools}
\usepackage{romannum}
\usepackage{graphicx}
\usepackage{enumitem}
\usepackage{colortbl}
\usepackage{amsmath,graphicx}
\usepackage{amsmath,amssymb,amsfonts}
\usepackage{algorithmic}
\usepackage{graphicx}
\usepackage{textcomp}
\usepackage{xcolor}
\usepackage{mathrsfs}
\graphicspath{{./fig/}}

\theoremstyle{break}

\newcommand{\ignore}[1]{ }

\newcommand{\figlbl}[1]{\label{fig.{#1}}}
\newcommand{\figref}[1]{Fig.~\ref{fig.{#1}}}

\newcommand{\beq}{\begin{equation}}
\newcommand{\eeq}{\end{equation}}

\setstcolor {red}

\usepackage{xifthen}
\newcounter{reviewer}
\newcounter{point}[reviewer]
\usepackage{bibentry}

\renewcommand{\thepoint}{Q\,\thereviewer.\arabic{point}} 

\newcommand{\shortreply}[2][]{\medskip \noindent \begin{sf}\textbf{Reply}:\  #2
	\ifthenelse{\equal{#1}{}}{}{ \hfill \footnotesize (#1)}%
	\medskip \end{sf}}
\title{
Long Polynomial Modular Multiplication using Low-Complexity Number Theoretic Transform\\
\thanks{This research was supported in
parts by the Semiconductor Research Corporation under contract number
2020-HW-2998, and by the National Science Foundation under grant number CCF-2243053.
}
}
\author{Sin-Wei Chiu,~\IEEEmembership{Student Member,~IEEE};
and Keshab K. Parhi,~\IEEEmembership{Fellow,~IEEE}
}

\begin{document}
\maketitle
\section{Scope}
\label{secsco} 
This tutorial aims to establish connections between polynomial modular multiplication over a ring to circular convolution and discrete Fourier transform (DFT). The main goal is to extend the well-known theory of DFT in signal processing (SP) to other applications involving polynomials in a ring such as homomorphic encryption (HE).
HE allows any third party to operate on the encrypted data without decrypting it in advance. 
Since most HE schemes are constructed from the ring-learning with errors (R-LWE) problem, efficient polynomial modular multiplication implementation becomes critical. Any improvement in the execution of these building blocks would have significant consequences for the global performance of HE. This lecture note describes three approaches to implementing long polynomial modular multiplication using the number theoretic transform (NTT): zero-padded convolution, without zero-padding, also referred to as negative wrapped convolution (NWC), and low-complexity NWC (LC-NWC).


\section{Relevance}
\label{secrel}

Homomorphic encryption (HE) schemes involve two fundamental operations: homomorphic multiplication and homomorphic addition. Most of the existing HE schemes are constructed from the R-LWE problem \cite{lyubashevsky2010ideal}. 
R-LWE-based HE schemes rely on polynomial multiplication/addition as the main building blocks, and the number of polynomial operations required grows with the multiplicative depth and width \cite{crockett2020low} of the desired function that needs to be homomorphically evaluated.
Since the ciphertexts of these schemes are in the form of polynomials, the addition and multiplication operations are performed on the polynomials. While the polynomial addition is simple (coefficient-wise modular addition), the polynomial modular multiplication is complex, especially when the degree of the polynomial is large and the word length of the coefficients is long. Therefore, the most time- and memory-consuming part of an R-LWE-based scheme is the long polynomial modular multiplication. Since polynomial multiplication can be viewed as a linear convolution of the coefficients, the intuitive way to compute the multiplication of two polynomials is to use the schoolbook algorithm with the time complexity of $O(n^2)$. However, the length of the polynomials, $n$, of a homomorphic encryption scheme can be in the range of thousands \cite{marcolla2022survey}. The time complexity of performing homomorphic multiplication can be reduced to $O(n\log n)$ using the number theoretic transform (NTT). 

In this paper, we provide a comprehensive guide toward efficient NTT-based polynomial modular multiplication. Three NTT-based approaches are described: zero-padded convolution, negative wrapped convolution (NWC), and low-complexity NWC (LC-NWC). Examples, derivations, and comparisons are presented. This tutorial is structured to provide an easy digest of the relatively complex topic.
\section{Prerequisites}
\label{secpr}

This article assumes only a familiarity
with discrete Fourier transform (DFT), fast Fourier transform (FFT), convolution, and basic polynomial operations.

\section{Problem Statement} 
\label{secpe}
Most HE schemes based on the R-LWE problem operate in the ring $R_{n,q} = \mathbb{Z}_{q}[x]/(x^n+1)$ \cite{marcolla2022survey}.
Polynomials over a ring $R_{n,q} = \mathbb{Z}_{q}[x]/(x^n+1)$ are defined as:
\begin{equation}
    p(x) = a_{0} + a_{1}x + \cdots + a_{n-2}x^{n-2} + a_{n-1}x^{n-1}
\end{equation}
where $n$ is a power-of-2 number. The coefficients are integers in $S = \{0,1,\ldots,q-1\}$. 
It is important to note that $n, q$ have a relation of $q \mod{2n} \equiv 1$. This ensures that the primitive 2$n$-th root of unity, $\psi_{2n}$, exists. The primitive 2$n$-th root of unity, $\psi_{2n}$, is also in set $S$, and $\psi^{n}_{2n} \equiv -1 ~(\text{mod} ~q)$, $\psi^{2n}_{2n} \equiv 1 ~(\text{mod} ~q)$. Let $\omega_{n}$ be the primitive $n$-th root of unity in $\mathbb{Z}_{q}$, which means $\omega^n_{n} \equiv 1 ~(\text{mod} ~q)$ and $\omega_{n} = \psi^2_{2n}$.

For example, let $n=4$, then we have a $3$rd order polynomial $p(x)$. Since we need to make sure $q \mod{2n} \equiv 1$, $q=17$ is selected. Next, let's find the 2$n$-th root of unity $\psi_{2n}$. If we try $\psi_{2n}=2$, we need to compute powers of $\psi_{2n}$ from $1$ to $2n=8$. We have $[2,4,8,16,32,64,128,256]$. Let's compute the modulo $q=17$ reduction of the elements in this vector, we have $[2,4,8,16,15,13,9,1]$. $\psi_{2n}^n$ should be $16$ ($-1 ~(\text{mod} ~q)$) and $\psi_{2n}^{2n}$ should be $1$ after modular reduction. Hence, $2$ is the 2$n$-th root of unity.


\subsection*{Modular polynomial multiplication}

In SP, the convolution operation is the most fundamental operation at the center of many developments related to the Fourier transform, superposition, impulse response, etc. It is well known that the convolution of two sequences $a$ and $b$ can be implemented using DFT. {Remember that with the DFT we can implement both circular and standard convolutions. For standard convolutions, it is necessary to zero-pad the two input sequences to ensure proper computation.}

{In these notes, we intend to provide a comprehensive explanation of the connection between polynomial modular multiplication, convolution, and DFT with polynomial multiplication in $\mathbb{Z}_{q}[x]/(x^n+1)$.}

For example, $a=[1~2]$ and $b=[1~-1]$. In MATLAB, the conv(a,b) commands
yields $[1~1~-2]$. To calculate the same results in the transform domain, we define $\Bar{a} = [1~2~0]$ and $\Bar{b} = [1~-1~0]$ as the zero-padded versions of $a$ and $b$; implement $DFT^{-1}(DFT(a) \odot DFT(b))$, where $\odot$ denotes the point-wise multiplication. This is the convolution theorem \cite{oppenheim2011discrete} in action! Similarly, we can use Z-transform and write $A(z) = 1 + 2z^{-1}$ and $B(z) = 1 - 1z^{-1}$, and compute $A(z)B(z)$ from which we can get the convolution result.

Assume that we have two polynomials $a(x)$ and $b(x)$ over the ring $R_{n,q} = \mathbb{Z}_{q}[x]/(x^n+1)$, where

\begin{align}
    a(x) &=  \sum_{j=0}^{n-1} a_{j}x^j \text{,}\\
    b(x) &=  \sum_{j=0}^{n-1} b_{j}x^j \text{.}
\end{align}

Remember that the coefficients of $a(x)$ and $b(x)$ have to be in the range of $[0, q-1]$. Let's assume that we want to compute the modular polynomial multiplication

\begin{equation}
    p(x) = a(x) \times b(x) \mod{(q, ~x^n+1)}
\end{equation}

It is important to point out that the operation $\text{mod}~(x^n+1)$ can be viewed as the negated mapping of conventional $\text{mod}~(x^n-1)$, i.e., the circular convolution \cite{blahut1983theory}. For example, $x^n \mod{x^n+1} = -1$ instead of $1$; $x^{n+1} \mod{x^{n}+1} = -x$ instead of $x$. In general, $x^{n+i} \mod{x^{n}+1} = -x^i$, where $i$ is an integer from $0$ to $n-1$. 

We can revisit the example of $a(x) = 1+2x$ and $b(x)=1-x$. Computing $a(x) \times b(x) \mod{(x^2-1)}$ is the same as computing the circular convolution. We have 
\begin{align*}
    &a(x) \times b(x) \mod{(x^2-1)} \\
    &= 1+x-2x^2 \mod{(x^2-1)}\\
    &= -1+x
\end{align*}
Computing $a(x) \times b(x) \mod{(x^2+1)}$ is the same as computing a negated circular convolution, commonly referred as negative wrapped convolution (NWC). We have 
\begin{align*}
    &a(x) \times b(x) \mod{(x^2+1)} \\
    &= 1+x-2x^2 \mod{(x^2+1)}\\
    &= 3+x
\end{align*}

The modular polynomial multiplication can be carried out using the convolution property \cite{oppenheim2011discrete} as:
\begin{align}
    \hat{p}(x) = INTT_{2n}(&NTT_{2n}(zeropadding(a(x)) \odot \nonumber\\
    &NTT_{2n}(zeropadding(b(x))),\label{eq:NTT_INTT}
\end{align}
\begin{equation}
    p(x) = \hat{p}(x) \mod{(q, ~x^n+1)}.
\end{equation}
The function $zeropadding(a(x))$ converts $a(x)$ from a length-$n$ polynomial to a length-$2n$ polynomial by padding $n$ zeros at the end.
\begin{equation}
    zeropadding(a(x)) = a(x) + \sum_{k=n}^{2n-1} a_{k}x^k, ~a_{k}=0 ~\forall~ k.
\end{equation}
where the NTT \cite{pedrouzo2017number}, a transformation similar to the DFT, is carried out in a finite ring \cite{mcclellan1979number}, where the twiddle factors are powers of an integer root of unity, i.e., $\omega^n_{n} \equiv 1 ~(\text{mod} ~q)$. Note that the twiddle factors in the DFT are expressed in terms of the complex exponential, $e^{-j2\pi/n}$, i.e., the $n$-th root of unity. The main reason that we are using NTT instead of conventional DFT is that the ciphertext in HE operates over integer arithmetic. No complex number calculations are required in NTT, unlike in DFT. Furthermore, DFT will introduce undesired additional errors in arithmetic operations due to truncation or rounding; these errors do not occur with NTT. NTT is defined as:
\begin{equation} \label{eq:NTT1}
    {A}_k = \sum_{j=0}^{n-1}  a_j\omega_{n}^{kj}\mod q, ~ k \in [0,n-1]
\end{equation}
We can represent NTT in a matrix form:
\begin{align}
    \mathbf{A} = \mathbf{W a}
\end{align}
where $\mathbf{A}$ and $\mathbf{a}$ are $n$-by-1 vectors, and $\mathbf{W}$ is the $n$-by-$n$ NTT matrix given by:
\begin{align}
    \begin{bmatrix}
        1 & 1 & 1 & \cdots & 1 \\
        1 & \omega & \omega^2 & \cdots & \omega^{n-1} \\
        1 & \omega^2 & \omega^4 & \cdots & \omega^{2(n-1)} \\
        \vdots & \vdots & \vdots & \ddots & \vdots \\
        1 & \omega^{n-1} & \omega^{2(n-1)} & \cdots & \omega^{(n-1)(n-1)}
    \end{bmatrix}
    \mod{q}
\end{align}
Note that $\mathbf{W}$ is a symmetric matrix. Let's assume we have $n=4$ and $q=17$. From the previous example, we know that $\psi_{2n}=2$ and $\omega = \psi^2_{2n}=4$. For these parameters, the $4$-by-$4$ NTT matrix is given by:
\begin{align*}
    \mathbf{W} =
    \begin{bmatrix}
        1 & 1 & 1 & 1 \\
        1 & 4 & 16 & 13 \\
        1 & 16 & 1 & 16 \\
        1 & 13 & 16 & 4 \\
    \end{bmatrix}
\end{align*}
Take $\mathbf{a}=[1~2~3~4]^T$ as an example, the output $\mathbf{A}$ from Equation (\ref{eq:NTT1}) before modular reduction will be $[10~109~100~91]^T$. After modular reduction, $\mathbf{A}$ will be $[10~7~15~6]^T$. 
INTT is defined as:
\begin{equation}
    {a}_j = n^{-1}\sum_{k=0}^{n-1}  A_k\omega_{n}^{-kj}\mod q, ~ j \in [0,n-1]
\end{equation}
Similar to NTT, we can also represent INTT in a matrix form:
\begin{align} \label{eq:INTT1}
    \mathbf{a} = \mathbf{W^{-1} A}
\end{align}
where $\mathbf{W^{-1}}$ is the inverse matrix of $\mathbf{W}$, and is given by:
\begin{align*}
    &\mathbf{W}^{-1} = \\ &n^{-1}
    \begin{bmatrix}
        1 & 1 & 1 & \cdots & 1 \\
        1 & \omega^{-1} & \omega^{-2} & \cdots & \omega^{-(n-1)} \\
        1 & \omega^{-2} & \omega^{-4} & \cdots & \omega^{-2(n-1)} \\
        \vdots & \vdots & \vdots & \ddots & \vdots \\
        1 & \omega^{-(n-1)} & \omega^{-2(n-1)} & \cdots & \omega^{-(n-1)(n-1)}
    \end{bmatrix}
    \text{ mod } q
\end{align*}
We can again create an example INTT matrix with the same parameters as above. First, we need to find $\omega^{-1}$ and $n^{-1}$. We can do so by finding the inverse of $\omega$ and $n$, i.e., $\omega \omega^{-1}~\text{mod}~{q} \equiv 1$ and $n n^{-1}~\text{mod}~{q} \equiv 1$. Therefore, for $\omega = 4$ and $n=4$, we have $\omega^{-1}=13$ and $n^{-1}=13$. The example INTT matrix is shown below:

\begin{align*}
    \mathbf{W}^{-1} = 13&
    \begin{bmatrix}
        1 & 1 & 1 & 1 \\
        1 & 13 & 16 & 4 \\
        1 & 16 & 1 & 16 \\
        1 & 4 & 16 & 13 \\
    \end{bmatrix}
    \text{mod } q \\
    =&
    \begin{bmatrix}
        13 & 13 & 13 & 13 \\
        13 & 16 & 4 & 1 \\
        13 & 4 & 13 & 4 \\
        13 & 1 & 4 & 16 \\
    \end{bmatrix}
\end{align*}
Take $\mathbf{A}=[10~7~15~6]^T$ from the previous example, the output $\mathbf{a}$ from Equation (\ref{eq:INTT1}) before modular reduction will be $[494~308~377~293]^T$. After modular reduction, $\mathbf{a}$ will be $[1~2~3~4]^T$, which is the same as what we started with.
Continuing from Equation (\ref{eq:NTT_INTT}),
we can compute $\hat{p}(x)$, the 2$n$-point polynomial multiplication output,
\begin{equation}
    \hat{p}(x) = \sum_{j=0}^{2n-1} \hat{p}_{j}x^j \text{, and } \hat{p}_{j} = \sum_{i=0}^{j} a_{i} b_{(j-i)}.
\end{equation}
The desired reduced product $p(x)$ can be calculated by using two 2$n$-point NTT and one 2$n$-point INTT followed by a modular polynomial reduction of $(q, x^n+1)$. \figref{polymul_BD}(a) shows a block diagram of modular polynomial multiplication using this approach.


Although we can correctly obtain the modular polynomial multiplication output, appending zeros to the original input polynomials to a length of 2$n$ and using an additional modular polynomial reduction block at the end are not efficient for computing the modular multiplication. Furthermore, the use of an $n$-point NTT instead of a $2n$ point NTT is desirable.

\section{Solution}
\label{secso}

We can use a different approach that does not require the $zeropadding()$ functions and the modular polynomial multiplication block. This approach is referred to as the negative wrapped convolution (NWC) \cite{lyubashevsky2008swifft}. Before we delve into the concept of NWC, it is crucial to point out that the conventional circular convolution cannot be applied to solve this problem since it requires modulo $(x^n+1)$ (negacyclic) operations instead of modulo $(x^n-1)$ (cyclic) operations.

\begin{figure}[h]
\centering
\resizebox{0.48\textwidth}{!}{
\includegraphics{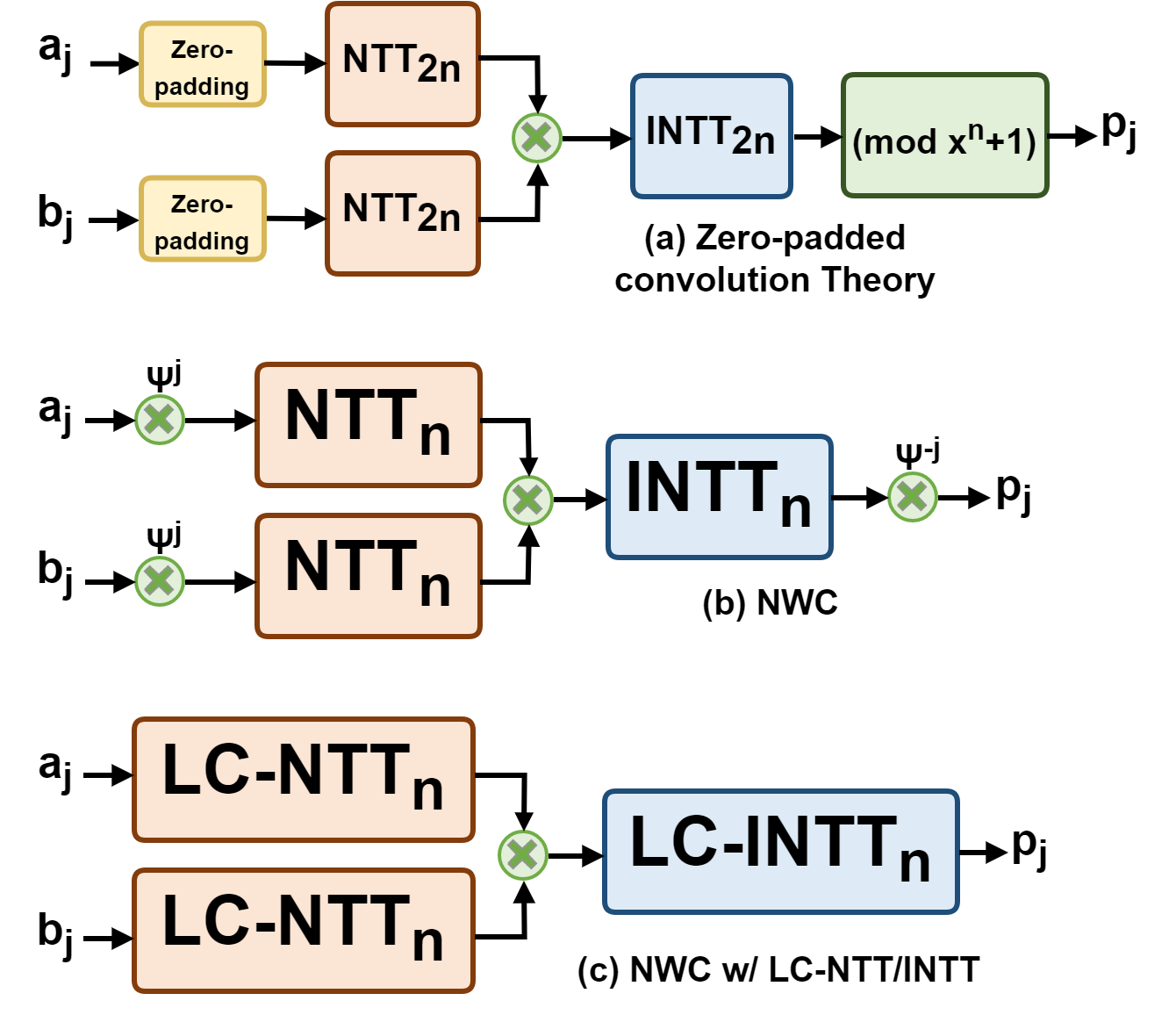}}
\caption{Block diagrams of modular polynomial multiplication (a) Zero-padded convolution theory. (b) NWC. (c) NWC with low-complexity NTT/INTT.}
\figlbl{polymul_BD}
\end{figure}

\subsection*{Negative wrapped convolution}

When computing polynomial multiplication $a(x) \times b(x) \mod{(q, ~x^n+1)}$ over the ring $R_{n,q} = \mathbb{Z}_{q}[x]/(x^n+1)$, NWC can be performed as:
\begin{equation}
    \widetilde{p}(x) = INTT_{n}(NTT_{n}(\widetilde{a}(x)) \odot NTT_{n}(\widetilde{b}(x))),
\end{equation}
\begin{equation}
    p(x) = \sum_{j=0}^{n-1}  \widetilde{p}_{j}  \psi_{2n}^{-j}x^j \mod q. \label{eq:ntt_post}
\end{equation}
where $\odot$ denotes point-wise multiplication and 
\begin{align}
    \widetilde{a}(x) &= \sum_{j=0}^{n-1} a_{j}  \psi_{2n}^{j}x^j \mod q, \label{eq:ntt_prea}\\
    \widetilde{b}(x) &= \sum_{j=0}^{n-1}  b_{j}  \psi_{2n}^{j}x^j \mod q. \label{eq:ntt_preb}
\end{align}

NWC makes sure that no zero-padding is required for the operation. With NWC, the desired reduced product $p(x)$ can be calculated by using two $n$-point NTT operations and one $n$-point INTT. Although we are able to reduce the 2n-point polynomial multiplication to a n-point polynomial multiplication, there are some tradeoffs. NWC requires pre-processing before NTT and post-processing after INTT. Equations (\ref{eq:ntt_prea}) and (\ref{eq:ntt_preb}) describe the pre-processing step where the input is multiplied by the 2$n$-th roots of unity raised to the power of $j$, $\psi_{2n}^j$, and Equation (\ref{eq:ntt_post}) represents the post-processing step where the output coefficients of INTT are multiplied by the inverse of 2$n$-th roots of unity raised to the power of $j$, $\psi_{2n}^{-j}$.
By combining the pre-processing and NTT, we have:
\begin{align}
    \mathbf{\widetilde{A}} = \mathbf{W\boldsymbol{\Psi} a} \label{eqn:NTT}
\end{align}
where $\boldsymbol{\Psi}$ is a $n$-by-$n$ diagonal matrix, whose diagonal terms are $\psi_{2n}^{j}$:
\begin{align}
    \begin{bmatrix}
        1 & 0 & 0 & \cdots & 0 \\
        0 & \psi_{2n}^{1} & 1 & \cdots & 0 \\
        0 & 0 & \psi_{2n}^{2} & \cdots & 0 \\
        \vdots & \vdots & \vdots & \ddots & \vdots \\
        0 & 0 & 0 & \cdots & \psi_{2n}^{n-1}
    \end{bmatrix}
    \mod{q}
\end{align}
We can also combine the post-processing with INTT:
\begin{align}
    \mathbf{a} = \boldsymbol{\Psi^{-1}} \mathbf{W^{-1} \widetilde{A}} \label{eqn:INTT}
\end{align}
where $\boldsymbol{\Psi^{-1}}$ is also a $n$-by-$n$ diagonal matrix, whose diagonal terms are $\psi_{2n}^{-j}$.

To prove the correctness of NWC, the coefficients of the NWC NTT outputs are denoted as:
\begin{align}
    \widetilde{A}_k &= \sum_{i=0}^{n-1} a_{i} \psi_{2n}^{i} \omega_{n}^{ki} 
\end{align}
\begin{align}
    \widetilde{B}_k &= \sum_{j=0}^{n-1} b_{j} \psi_{2n}^{j} \omega_{n}^{kj} 
\end{align}
The point-wise multiplication of two NWC NTT outputs is given by:
\begin{align}
    \widetilde{P}_k = \widetilde{A}_k  \widetilde{B}_k = \sum_{i=0}^{n-1} \sum_{j=0}^{n-1} a_{i} b_{j} \psi_{2n}^{(i+j)} \omega_{n}^{k(i+j)}
\end{align}
Applying NWC INTT to $\widetilde{P}_k$, we have:
\begin{align}
    {p}_l &= n^{-1} \psi^{-l}_{2n} \left( \sum_{k=0}^{n-1} \left(\sum_{i=0}^{n-1} \sum_{j=0}^{n-1} a_{i} b_{j} \psi_{2n}^{(i+j)} \omega_{n}^{k(i+j)} \right) \omega_{n}^{-lk} \right) \nonumber \\ 
    &= n^{-1} \psi^{-l}_{2n} \sum_{i=0}^{n-1} \sum_{j=0}^{n-1} a_{i} b_{j} \psi_{2n}^{(i+j)} \sum_{k=0}^{n-1} \omega_{n}^{k(i+j-l)}  \label{eq:NWCINTT1}
\end{align}
Since 
\[\sum_{k=0}^{n-1} \omega_{n}^{k(i+j-l)} =
\begin{cases}
    n & \text{if } (i+j-l) = n \text{ or } 0\\
    0 & \text{otherwise}
\end{cases}
\]
Equation (\ref{eq:NWCINTT1}) can be expressed as: 
\begin{align}
    {p}_l &= \psi^{-l}_{2n} \sum_{i=0}^{l} a_{i} b_{l-i} \psi_{2n}^{l} + \psi^{-l}_{2n} \sum_{i=l+1}^{n-1} a_{i} b_{n+l-i} \psi_{2n}^{l+n} \nonumber \\
    &= \sum_{i=0}^{l} a_{i} b_{l-i} + \sum_{i=l+1}^{n-1} a_{i} b_{n+l-i} \psi_{2n}^{n} \label{eq:NWCINTT2} \\
    &= \sum_{i=0}^{l} a_{i} b_{l-i} - \sum_{i=l+1}^{n-1} a_{i} b_{n+l-i} \label{eq:NWCINTT3}
\end{align}

Note that subtraction of the second term in Equation (\ref{eq:NWCINTT3}) implicitly carries out the polynomial modulo $ (x^n + 1) $. The negative term results from the fact that $\psi_{2n}^{n}=-1$ in Equation (\ref{eq:NWCINTT2}). We can also connect Equation (\ref{eq:NWCINTT2}) to the circular convolution. If we remove the $\psi$ from the equation, Equation (\ref{eq:NWCINTT3}) becomes:
\begin{equation}
    \sum_{i=0}^{l} a_{i} b_{l-i} + \sum_{i=l+1}^{n-1} a_{i} b_{n+l-i}
    \label{eqn:pos_wrap}
\end{equation}
It is easy to see that this is exactly a circular convolution. It is also called positive wrapped convolution because of the plus term.

\figref{polymul_BD}(b) shows a block diagram of modular polynomial multiplication using NWC. The NWC algorithm is described in Algorithm \ref{algmpoly}.


\begin{algorithm}[htbp]
\caption{\textbf{Negative Wrapped Convolution~\cite{lyubashevsky2008swifft}}}
\label{algmpoly}
\hspace*{\algorithmicindent} \textbf{Input:} $a(x), b(x) \in R_{n,q}$

\hspace*{\algorithmicindent} \textbf{Output:} ${p}(x) = a(x) \times b(x) \mod (x^n+1,q)$
 
\begin{algorithmic}[1]
     \STATE $\widetilde{a}(x) =\sum_{j=0}^{n-1} a_j\psi_{2n}^jx^j \mod q$ \\
     $\widetilde {b}(x) = \sum_{j = 0}^{n-1} b_j\psi_{2n}^jx^j \mod q$
     \STATE $\widetilde{A}(x): \widetilde{A}_k = \sum_{j=0}^{n-1} \widetilde a_j\omega_{n}^{kj}\mod q$, $ k \in [0,n-1]$ \\ 
     $\widetilde{B}(x): \widetilde{B}_k = \sum_{j=0}^{n-1} \widetilde b_j\omega_{n}^{kj}\mod q$, $ k \in [0,n-1]$ 
     \STATE  $\widetilde{P}(x) = \widetilde{A}(x) \odot \widetilde{B}(x) = \sum_{k=0}^{n-1} \widetilde {A}_k \widetilde{B}_kx^k$ 
     \STATE $\tilde{p}(x): \widetilde{p}_j = n^{-1}\sum_{k=0}^{n-1} \widetilde{P}_k\omega_{n}^{-kj}\mod q$, $j \in [0,n-1]$
     \STATE  $p(x) =  \sum_{j=0}^{n-1}  \widetilde{p_j}  \psi_{2n}^{-j}x^j \mod q$
\end{algorithmic}
\end{algorithm}

Although we can reduce the length-2$n$ polynomial multiplication to length-$n$ by using NWC, there are still some tradeoffs. Additional weighted operations are required before NTT and after INTT. This requires a total of 2$n$ additional large coefficient modular multiplications compared to classic NTT/INTT computation. Recent works \cite{zhang2020highly,roy2014compact} have presented a new method to merge the weighted operations into the butterfly operations. This method is able to merge the pre-processing portion into the NTT block with low-complexity NTT and the post-processing portion into the INTT block with low-complexity INTT. This is illustrated in the block diagram shown in \figref{polymul_BD}(c).




\subsection*{Low-complexity NTT}

The low-complexity NTT merges the weighted operation before NTT in Step 2 of Algorithm \ref{algmpoly} by changing the twiddle factors. In particular, the new NTT operation is re-represented as $ \widetilde{A}_k$ and $ \widetilde{A}_{k+n/2}$ by using the decimation-in-time (DIT) method \cite{duhamel1990fast} in FFT. This method divides the input sequence into the sequence of even and odd numbered samples. Thus, the name ``decimation-in-time".

The NTT equation for NWC is described by:
\begin{align} \label{eq:1}
    \widetilde{A}_k = \sum_{j=0}^{n-1} a_j \psi_{2n}^{j} \omega_{n}^{kj}\mod q,
\end{align}
we can rewrite Equation (\ref{eq:1}) by splitting the summation into two groups: one containing the even and the other containing odd coefficients. For $k=0,1,...,n-1$:
\begin{align*}
    \widetilde{A}_k &= \sum_{j=0}^{n/2-1} a_{2j} \psi_{2n}^{2j} \omega_{n}^{2kj} \\
    &+ \sum_{j=0}^{n/2-1} a_{(2j+1)} \psi_{2n}^{2j+1} \omega_{n}^{k(2j+1)} \mod{q}
\end{align*}
With the scaling property of twiddle factors, $\omega^{k/m}_{n/m}=\omega^{k}_{n}$:
\begin{align*}
    \widetilde{A}_k &= \sum_{j=0}^{n/2-1} a_{2j} \psi_{n}^{j} \omega_{n/2}^{kj}\\
    & + \psi_{2n}\omega^{k}_{n} \sum_{j=0}^{n/2-1} a_{(2j+1)} \psi_{n}^{j} \omega_{n/2}^{kj} \mod{q}
\end{align*}
Then we can group them into two parts based on the size of the index $k$. For indices $k>n/2-1$, we rewrite them as $k+n/2$, where $k=0,1,...,n/2-1$ By applying the symmetry property of twiddle factors $(\omega^{k+n/2}_{n}=-\omega^{k}_{n})$ and the periodicity property of twiddle factors $(\omega^{k+n}_{n}=\omega^{k}_{n})$, we have:
\begin{align*}
    \widetilde{A}_k &= a^{(0)}_k + \psi_{2n} \omega^k_n a_k^{(1)} \mod q , \\
    \widetilde{A}_{k+n/2} &= a^{(0)}_k - \psi_{2n} \omega^k_n a_k^{(1)} \mod q ,
\end{align*}
where $k\in [0,\frac{n}{2}-1]$ and 
\begin{align}
    a^{(0)}_k &=  \sum_{j=0}^{n/2-1}  a_{2j}\psi^j_{n} \omega_{n/2}^{kj}\mod q,\\
    a^{(1)}_k &= \sum_{j=0}^{n/2-1}  a_{(2j+1)}\psi^j_{n} \omega_{n/2}^{kj}\mod q.
\end{align}

It is easy to see that $a^{(0)}_k$ and $a^{(1)}_k$ are essentially same as Equation (\ref{eq:1}); the only difference is that they are scaled down to $n/2$ points. By recursively applying the decimation process to $a^{(0)}_k$ and $a^{(1)}_k$ to 2-point NTT, we can get the structure shown in Figure \figref{LC-NWC8} (upper left). Also, Since $\omega_{n} = \psi_{2n}^2  \mod q$, the integers $\psi_{2n}$ and $\omega_{n}^{k}$ can be merged to an integer, $\psi_{2n} \omega^k_n = \psi^{(2k+1)}_{2n}$. Thus,

\begin{align*}
    \widetilde{A}_k &= a^{(0)}_k + \psi^{(2k+1)}_{2n} a_k^{(1)} \mod q , \\
    \widetilde{A}_{k+n/2} &= a^{(0)}_k - \psi^{(2k+1)}_{2n} a_k^{(1)} \mod q .
\end{align*}

We can further represent this architecture in a matrix form.
\begin{align}
    \mathbf{\widetilde{A}} &=\mathbf{W\boldsymbol{\Psi} a} \nonumber \\
    &= \mathbf{\widetilde{W} a} \label{eq:LCNWCNTT}
\end{align}
where $\mathbf{\widetilde{W}}$ is a modified version NTT matrix for NWC:
\begin{align}
\mathbf{\widetilde{W}} = 
    \begin{bmatrix}
        1 & \psi & \psi^2 & \cdots & \psi^{n-1} \\
        1 & \psi \omega & \psi^2 \omega^2 & \cdots & \psi^{n-1} \omega^{n-1} \\
        1 & \psi \omega^2 & \psi^2 \omega^4 & \cdots & \psi^{n-1} \omega^{2(n-1)} \\
        \vdots & \vdots & \vdots & \ddots & \vdots \\
        1 & \psi \omega^{(n-1)} & \psi^2 \omega^{2(n-1)} & \cdots & \psi^{n-1} \omega^{(n-1)(n-1)} \\
    \end{bmatrix}
    \nonumber
\end{align}
Note that the $\mathbf{\widetilde{W}}$ matrix is not a symmetric matrix like $\mathbf{W}$.

\begin{figure*}[t!]
\centering
\resizebox{0.98\textwidth}{!}{
\includegraphics{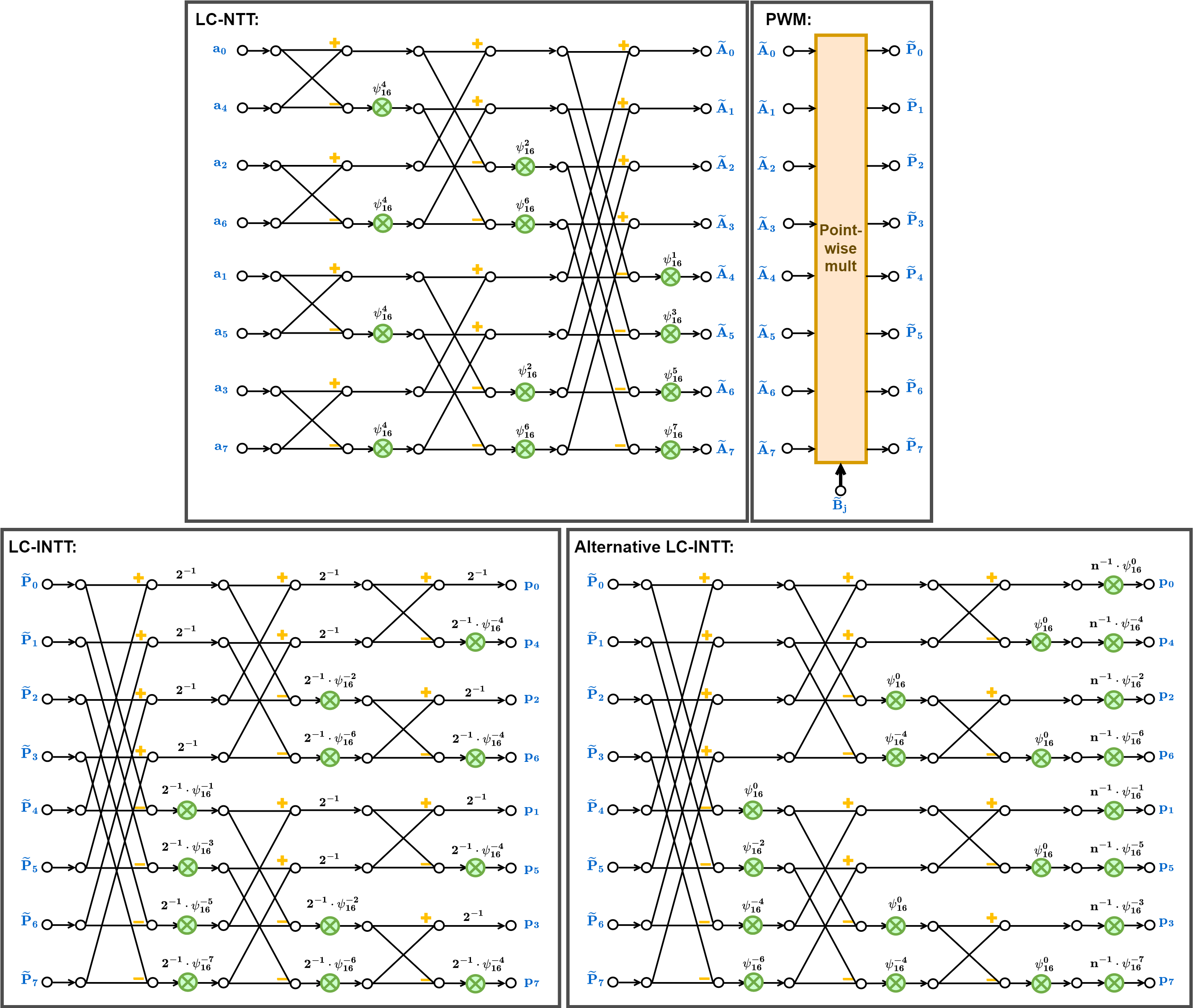}}
\caption{The data flow graph of an 8-point low-complexity negative wrapped convolution.}
\figlbl{LC-NWC8}
\end{figure*}

\subsection*{Low-complexity INTT}
The improved INTT algorithm merges not only the weighted operation but also the multiplication with constant $n^{-1}$ into the butterfly operations, as presented in~\cite{zhang2020highly}.

The low-complexity NWC operation can be described as:

\begin{align}
    \mathbf{p} &= \boldsymbol\Psi^{-1} \mathbf{W^{-1} \widetilde{P}} \nonumber \\
    &= \mathbf{\widetilde{W}}^{-1} \mathbf{\widetilde{P}} \label{eq:LCNWCINTT}
\end{align}
and
\begin{align}
\mathbf{\widetilde{W}}^{-1} &= \\ 
    & n^{-1}
    \begin{bmatrix}
        1 & 1 & \cdots & 1 \\
        \psi^{-1} & \psi^{-1} \omega^{-1} & \cdots & \psi^{-1} \omega^{-(n-1)}\\
        \psi^{-2} & \psi^{-2} \omega^{-2} & \cdots & \psi^{-2} \omega^{-2(n-1)}\\
        \vdots & \vdots & \ddots & \vdots \\
        \psi^{-(n-1)} & (\psi \omega)^{-(n-1)} & \cdots & (\psi \omega^{(n-1)})^{-(n-1)}
    \end{bmatrix}
    \nonumber
\end{align}
where $\mathbf{\widetilde{W}}^{-1}$ is the inverse of $\mathbf{\widetilde{W}}$.
Equation (\ref{eq:LCNWCINTT}) can be interpreted as the transpose of the low-complexity NTT followed by $n^{-1}$ scaling. We can obtain the low-complexity INTT structure by first transposing the low-complexity NTT structure, changing the twiddle factors to its inverse, and adding multiply by $2^{-1}$ at the end of every stage, which is equivalent to multiplying $n^{-1}$ $(n^{-1} = (2^{-1})^{(\log_{2}n)})$. Thus, transposing the NTT structure in \figref{LC-NWC8} (upper left), replacing the twiddle factors by their inverse, and inserting $2^{-1}$ after every stage leads to the low-complexity INTT structure in \figref{LC-NWC8} (lower left).

Although we could derive the structure based on intuition, we could still derive the low-complexity INTT based on the decimation-in-frequency (DIF) method \cite{duhamel1990fast} in FFT. This method divides the output sequence into the sequence of even and odd numbered samples. Thus, the name ``decimation-in-frequency”.
The INTT equation for negative wrapped convolution is given by:
\begin{align} \label{eq:2}
    {p}_k = n^{-1}\psi^{-k}_{2n}\sum_{j=0}^{n-1} \widetilde{P}_j\omega_{n}^{-kj}\mod q
\end{align}
we can rewrite Equation (\ref{eq:2}) by splitting the items in the summation into two parts according to the size of the index of $\widetilde{P}_j$. For $k=0,1,...,n-1$:
\begin{equation*}
    p_k = n^{-1} \psi^{-k}_{2n} \left( \sum_{j=0}^{n/2-1} \widetilde{P}_j \omega_{n}^{-kj} + \sum_{j=n/2}^{n-1} \widetilde{P}_j \omega_{n}^{-kj} \right) \mod{q}
\end{equation*}
Based on the symmetry property and periodicity property of twiddle factors, the index of the second sum can be changed from $[n/2, n-1]$ to $[0, n/2-1]$:
\begin{align*}
    p_k &= n^{-1} \psi^{-k}_{2n} \left[ \sum_{j=0}^{n/2-1} \widetilde{P}_j \omega_{n}^{-kj} \right. \\ 
    &+ \left. \sum_{j=0}^{n/2-1} \widetilde{P}_{(j+n/2)} \omega_{n}^{-k(j+n/2)} \right] \mod{q}
\end{align*}
\begin{align*}
    &= n^{-1} \psi^{-k}_{2n} \left[ \sum_{j=0}^{n/2-1} \widetilde{P}_j \omega_{n}^{-kj} \right. \\
    &+ \left. (-1)^{k} \sum_{j=0}^{n/2-1} \widetilde{P}_{(j+n/2)} \omega_{n}^{-kj} \right] \mod{q}
\end{align*}
According to the parity of k, we can group them into two parts, where $k=0,1,...,n/2-1$:
\begin{align*}
    p_{2k} &= n^{-1} \psi^{-2k}_{2n} \left[ \sum_{j=0}^{n/2-1} \widetilde{P}_j \omega_{n}^{-2kj} \right. \\
    &+ \left. (-1)^{2k} \sum_{j=0}^{n/2-1} \widetilde{P}_{(j+n/2)} \omega_{n}^{-2kj} \right] \mod{q} 
\end{align*}
\begin{align*}
    p_{2k+1} &= n^{-1} \psi^{-(2k+1)}_{2n} \left[ \sum_{j=0}^{n/2-1} \widetilde{P}_j \omega_{n}^{-(2k+1)j} \right. \\
    &+ \left. (-1)^{(2k+1)} \sum_{j=0}^{n/2-1} \widetilde{P}_{(j+n/2)} \omega_{n}^{-(2k+1)j} \right] \mod{q}
\end{align*}
With the scaling property of twiddle factors, we can simplify the equations as:
\begin{align*}
    &p_{2k} = (\frac{n}{2})^{-1} \psi^{-k}_{n}  \sum_{j=0}^{n/2-1} \left[ \frac{\widetilde{P}_j+\widetilde{P}_{(j+n/2)}}{2} \right] \omega_{n/2}^{-kj} \mod{q} \\
\end{align*}
\begin{align*}
    &p_{2k+1} = \\ &(\frac{n}{2})^{-1} \psi^{-k}_{n}  \sum_{j=0}^{n/2-1} \left\{\left[ \frac{\widetilde{P}_j-\widetilde{P}_{(j+n/2)}}{2}\right] \psi^{-1}_{2n} \omega_{n}^{-j} \right\} \omega_{n/2}^{-kj}\\
    &\mod{q}
\end{align*}
Let
\begin{align}
    \widetilde{P}^{(0)}_j &= \frac{\widetilde{P}_j+\widetilde{P}_{j+n/2}}{2} \mod q, \nonumber \\
    \widetilde{P}^{(1)}_j &= \frac{\widetilde{P}_j-\widetilde{P}_{j+n/2}}{2} \psi_{2n}^{-1} \omega_n^{-j} \mod q. \nonumber
\end{align}
We have
\begin{align}
    p_{2k} &= (\frac{n}{2})^{-1}  \psi^{-k}_{n} \sum_{j=0}^{n/2-1}  \widetilde{P}^{(0)}_j \omega^{-kj}_{n/2} \mod q,
\end{align}
\begin{align}
    p_{2k+1} &= (\frac{n}{2})^{-1} \psi^{-k}_{n}\sum_{j=0}^{n/2-1}   \widetilde{P}^{(1)}_j \omega^{-kj}_{n/2} \mod q .
\end{align}

Similar to NTT, we can easily see that $p_{2k}$ and $p_{2k+1}$ are essentially the same as Equation (\ref{eq:2}) except scaled down to $n/2$ points. By recursively applying the decimation process to $p_{2k}$ and $p_{2k+1}$ to 2-point NTT, we can get the structure shown in Figure \figref{LC-NWC8} (lower left). Note that when $n=2$, $(\frac{n}{2})^{-1} = 1$, $\psi^{-k}_{n} = 1$, and also $\omega^{-kj}_{n/2} = 1$. In addition, the integers $\psi^{-1}_{2n}$ and $\omega^{-j}_n$ can be merged to an integer, $\psi^{-1}_{2n} \omega^{-j}_n = \psi^{-(2j+1)}_{2n}$.
The data flow graph of the entire 8-point low-complexity negative wrapped convolution is shown in \figref{LC-NWC8}.


Unlike the NTT butterfly architecture, the intermediate results after the modular addition and modular subtraction operations in the INTT butterfly need to be multiplied by $2^{-1}\mod{q}$. Although it seems like this will add additional multipliers to the INTT block, the modular multiplication by $2^{-1}$ can be implemented without a modular multiplier. 
\begin{align}
    \frac{x}{2} \mod{q} = 
    \begin{cases}
        \frac{x}{2} & \text{if } x \text{ is even} \\
        \lfloor \frac{x}{2} \rfloor + \frac{q+1}{2} \mod{q}& \text{if } x \text{ is odd} \label{eqn:two_inv_sel}
    \end{cases}
\end{align}

If $x$ is even, $x \times 2^{-1}$ can be implemented as a right shift operation, i.e., $x \gg 1$. Here, $\lfloor~\rfloor$ is the floor function that maps a number to the closest integer that is smaller than or equal to the number. The $\gg$ operation can be implemented easily in hardware. For example, a right shift by $1$ bit operation on $8$ ($100$ in binary) results in 4 ($010$ in binary). 

If $x$ is odd, $x \times 2^{-1}$ can be represented as:

\begin{align}
    \frac{x}{2} &\equiv (2 \lfloor \frac{x}{2}\rfloor + 1)\frac{q+1}{2} \mod{q} \label{eq:two_inv_org} \\
                &\equiv  \lfloor \frac{x}{2} \rfloor (q+1) + \frac{q+1}{2} \mod{q} \nonumber \\
                &\equiv  \lfloor \frac{x}{2} \rfloor + \frac{q+1}{2} \mod{q} \label{eq:two_inv}
\end{align}

The term $(2 \lfloor \frac{x}{2}\rfloor + 1)$ in Equation (\ref{eq:two_inv_org}) is equivalent to an odd number $x$; the term $(\frac{q+1}{2} \mod{q})$ is equivalent to $2^{-1} \mod{q}$ since $(\frac{q+1}{2} \times 2 \mod{q}) \equiv 1 \text{ mod } q$. $\lfloor \frac{x}{2} \rfloor$ can be implemented as $(x \gg 1)$, and $(q+1)/2$ is a constant. Hence, no modular multiplications are required. This operation requires one modular adder and a multiplexer. Here, the multiplexer is used to select one of the two options in Equation (\ref{eqn:two_inv_sel}) as output depends on whether the input is even or odd.

\subsection*{Alternative Low-complexity INTT }

There is another straightforward way of constructing an alternative low-complexity INTT structure (Alt-LC-INTT). We can merge the two post-processing multipliers $n^{-1}$ and $\psi_{2n}^j$ into a single equivalent multiplier. The structure is shown in \figref{LC-NWC8} (lower right). While the number of multipliers seems to be larger than the standard LC-INTT, the obvious advantage of this structure is that no shifting operations are required in each butterfly unit. More comparisons will be discussed in Section \ref{secwhat}.

\section{Numerical Example}
\label{secne}

\begin{figure*}[htbp!]
\centering
\resizebox{\textwidth}{!}{
\includegraphics{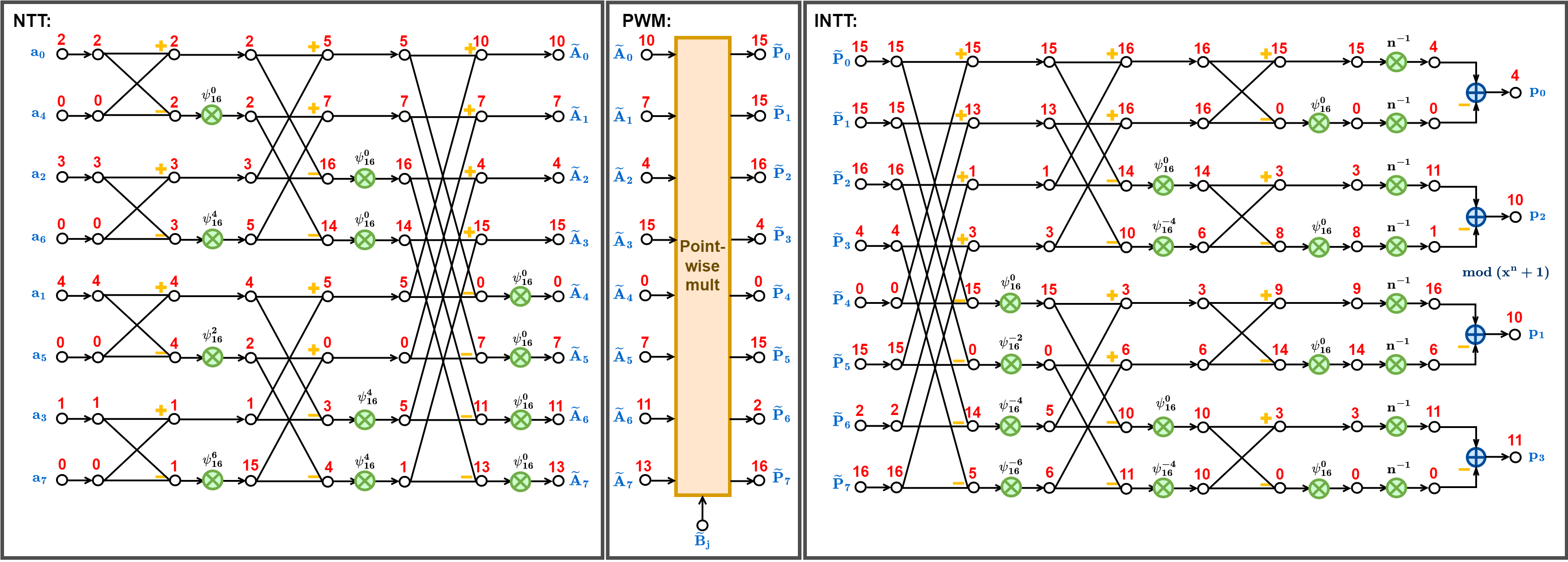}}
\caption{An example of degree 4 modular polynomial multiplication using 8-point convolution.}
\figlbl{Conv_Ex}
\end{figure*}

\figref{Conv_Ex} illustrates an example of length-4 modular polynomial multiplication using zero padding and 8-point NTT. In this example, $n=8$, $q = 17$, where $q \mod{2n} \equiv 1$ and it is also a prime, $n^{-1}=15$ $(8 \times 15 = 120 \equiv 1 \mod 17)$. Since $\psi^n_{2n} \mod{q} \equiv -1$, we can select $\psi_{16}=3$ $(3^8=6561 \equiv -1 \mod{17})$, and $\psi^{-1}_{16} = 6$ $(3 \times 6 = 18 \equiv 1 \mod{17})$. Assume that both $a(x)$ and $b(x)$ are $x^3+3x^2+4x+2$. To begin the computation, we need to first pad 4 zeros to the inputs, and then feed the inputs to the NTT block. After the NTT block, we will perform point-wise multiplications. Since we assume $a(x)$ and $b(x)$ are the same, the coefficients of the results of point-wise multiplications $\widetilde{P}_j$ will be $[10^2, 7^2, 4^2, 15^2, 0^2, 7^2, 11^2, 13^2] \mod{17}$ that is equivalent to $[15, 15, 16, 4, 0, 15, 2, 16]$. Next, we feed those outputs from point-wise multiplications to the INTT block. The INTT block is similar to the NTT block with only two differences. One, the 2$n$-th roots are now replaced with the inverse of 2$n$-th roots. Two, additional multipliers are added for multiplying $n^{-1}$. The INTT block outputs 8 coefficients. Since we are computing modular polynomial multiplication  $\mod (x^4+1, 17)$, the convolution result $x^6+6x^5+11x^3+11x^2+16x+4 \mod{(x^4+1, 17)}$ becomes $11x^3+(11-1)x^2+(16-6)x+(4-0) \mod{17} \equiv 11x^3+10x^2+10x+4$.

\figref{NWC_Ex} illustrates simple examples of NTT and INTT for negative wrapped convolution. On the left is NWC with classic NTT/INTT. Let's consider $n = 4$, $q = 17$, and $n^{-1}=13$ $(4 \times 13 = 52 \equiv 1 \mod 17)$ Since $\psi^n_{2n} \mod{q} \equiv -1$, we can select $\psi_{8}=2$ $(2^4=16 \equiv -1 \mod{17})$, and $\psi^{-1}_{8} = 9$ $(2 \times 9 = 18 \equiv 1 \mod{17})$. Let's consider the same example that both $a(x)$ and $b(x)$ are $(x^3+3x^2+4x+2)$. The first step of NWC is NTT with preprocessing, which correspond to steps 1 and 2 of Algorithm \ref{algmpoly}. We multiply each coefficient of $a(x)$ with the 2$n$-th root to the power of its exponent; this gives us the weighted $\widetilde{a}(x)$. After we obtain $\widetilde{a}(x)$, we feed the weighted input into NTT. Note that twiddle factor $\omega$ is the $n$-th root of unity, which means $\psi^2 = \omega$. 

Step 3 of Algorithm \ref{algmpoly} takes the outputs of both NTT blocks and performs point-wise multiplication. Since we assume $a(x)$ and $b(x)$ are the same, the coefficients of the results of point-wise multiplications $\widetilde{P}_j$ will be $[13^2, 7^2, 15^2, 7^2] \mod{q}$ that is equivalent to $[16, 15, 4, 15]$.

Steps 4 and 5 of Algorithm \ref{algmpoly} are feeding $\widetilde{P}(x)$ into the weighted INTT block. 
Note that after the INTT block, there are weighted operations that multiply each coefficient with the inverse of 2$n$-th root to the power of its exponent. The polynomial we obtain at the output is $11x^3+10x^2+10x+4$. We can verify this result by computing $(x^3+3x^2+4x+2)^2 \mod{(x^4+1, 17)} = x^6+6x^5+17x^4+28x^3+28x^2+16x+4 \mod{(x^4+1, 17)}$, which is equivalent to $11x^3+10x^2+10x+4$.



\begin{figure*}[htbp!]
\centering
\resizebox{\textwidth}{!}{
\includegraphics{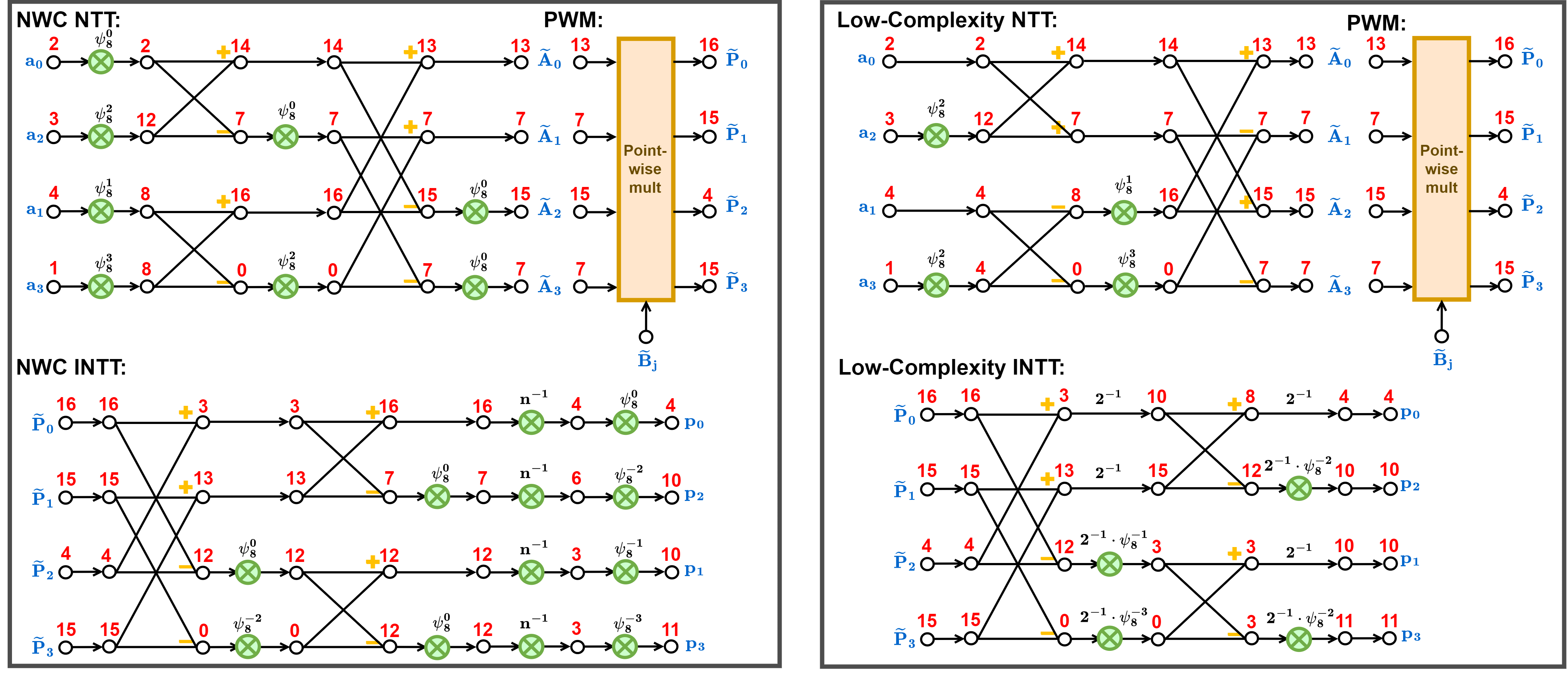}}
\caption{An example of 4-point negative wrapped convolution.}
\figlbl{NWC_Ex}
\end{figure*}

On the right of \figref{NWC_Ex} is an example of negative wrapped convolution with low-complexity NTT/INTT. We consider the same inputs as in \figref{NWC_Ex} that both $a(x)$ and $b(x)$ are $x^3+3x^2+4x+2$. $n = 4$, $q = 17$, $\psi = 2$, $\psi^{-1} = 9$ $(2 \times 9 = 18 \equiv 1 \mod{17})$, and $2^{-1}=9$ $(2 \times 9 = 18 \equiv 1 \mod{17})$. For the low-complexity NTT, the multipliers are now moved before the butterfly addition and subtraction. The output polynomial is the same as what we obtained from the previous example. For the standard low-complexity INTT, additional ``multiplication" of $2^{-1}$ is required after every butterfly addition and subtraction. However, no additional multipliers are implemented according to Equation (\ref{eq:two_inv}). The output polynomial is $11x^3+10x^2+10x+4$, the same as what we obtained from the previous example.
\section{What we have learned} \label{secwhat}

\subsection*{Comparisons}
Table \ref{table:mult_comp} compares the number of multipliers implemented in each method. The first method, zero padding and a polynomial modular reduction, requires $n\log 2n$ and $n\log 2n + 2n$ modular multiplications for NTT and INTT, respectively. The second method, the negative wrapped convolution, requires $\frac{n}{2}\log n + n$ and $\frac{n}{2}\log n + 2n$ modular multiplications for NTT and INTT, respectively. The improvement comes from reducing $2n$-point NTT/INTT to $n$-point NTT/INTT. However, the tradeoffs require adding $n$ multipliers to both NTT/INTT blocks. Last but not least, using the low-complexity NTT/INTT, we are able to remove the additional $n$ multipliers for NTT, and the additional $2n$ multipliers for INTT.

Table \ref{table:mult_comp} includes multipliers that multiply by $\psi_{2n}^{0} = 1$. If we remove those multipliers, the zero-padded convolution method requires $(n\log_2 n) - n + 1$ and $(n\log_2n) + n + 1$ modular multiplications for NTT and INTT, respectively. The NWC method requires $\frac{n}{2}\log n$ and $\frac{n}{2}\log n + n$ modular multiplications for NTT and INTT, respectively. The LC-NWC method requires the same number of modular multiplications for NTT and further reduces the number of modular multiplications for INTT by $n$. The comparison after excluding the multipliers by $1$ is shown in Table \ref{table:mult_comp_excld1}.

\begin{table}[h!]
\centering
\caption{The comparison of the numbers of multipliers}
\renewcommand{\arraystretch}{1.25}
\begin{tabular}{|c | c | c |} 
 \hline
  & \multicolumn{2}{c|}{$\#$ Multipliers} \\
 \cline{2-3} 
  & NTT & INTT \\
 \hline
 Conv. w/ mod  & $n\log 2n$ & $n\log 2n + 2n$\\ 
 \hline
 NWC & $\frac{n}{2}\log n + n$ & $\frac{n}{2}\log n + 2n$\\
 \hline
 LC-NWC & $\frac{n}{2}\log n$ & $\frac{n}{2}\log n$\\
 \hline
\end{tabular}
\label{table:mult_comp}
\end{table}

\begin{table}[h!]
\centering
\caption{The comparison of the numbers of multipliers excluding multiplication by 1}
\renewcommand{\arraystretch}{1.25}
\begin{tabular}{|c | c | c |} 
 \hline
  & \multicolumn{2}{c|}{$\#$ Multipliers} \\
 \cline{2-3} 
  & NTT & INTT \\
 \hline
 Conv. w/ mod  & $(n\log_2 n) - n + 1$ & $(n\log_2n) + n + 1$\\ 
 \hline
 NWC & $\frac{n}{2}\log n$ & $\frac{n}{2}\log n + n$\\
 \hline
 LC-NWC & $\frac{n}{2}\log n$ & $\frac{n}{2}\log n$\\
 \hline
\end{tabular}
\label{table:mult_comp_excld1}
\end{table}




Table \ref{table:save} illustrates how many modular multipliers can be saved by implementing the NWC methods compared to the traditional convolution theory for $n=\{1024, 2048, 4096\}$. Generally, implementing NWC will save about 46$\%$ for both NTT and INTT. If we implement NWC using low-complexity NTT/INTT, these numbers will go up to about 54$\%$ for NTT and about 60$\%$ for INTT.

Table \ref{table:save_excld1} illustrates how many modular multipliers can be saved by implementing the NWC methods compared to the traditional convolution theory for $n=\{1024, 2048, 4096\}$, excluding multiplication by $1$. Generally, implementing NWC will save about 45$\%$ and 46$\%$ for both NTT and INTT, respectively. If we implement NWC using low-complexity NTT/INTT, the percentages saved stay the same for NTT, but the percentages go up to about 54$\%$ for INTT.

\begin{table}[h!]
\centering
\caption{Percentage of the number of multipliers saved compared to the zero-padded convolution method}
\renewcommand{\arraystretch}{1.25}
\begin{tabular}{|c | c | c | c | c | c | c |}
 \hline
 \multirow{3 }{*}{n} & \multicolumn{4}{c|}{Percentage of $\#$ multipliers saved} \\
  \cline{2-5}
  & \multicolumn{2}{c|}{NWC} & \multicolumn{2}{c|}{LC-NWC} \\
  \cline{2-5}
  & NTT & INTT & NTT & INTT \\
 \hline
 1024 & 45.5 & 46.2 & 54.5 & 61.5 \\
 \hline
 2048 & 45.8 & 46.4 & 54.2 & 60.7 \\
 \hline
 4096 & 46.2 & 46.7 & 53.9 & 60.0 \\
 \hline
\end{tabular}
\label{table:save}
\end{table}
 
\begin{table}[h!]
\centering
\caption{Percentage of the number of multipliers saved compared to the zero-padded convolution method excluding multiplication by 1}
\renewcommand{\arraystretch}{1.25}
\begin{tabular}{|c | c | c | c | c | c | c |}
 \hline
 \multirow{3 }{*}{n} & \multicolumn{4}{c|}{Percentage of $\#$ multipliers saved} \\
  \cline{2-5}
  & \multicolumn{2}{c|}{NWC} & \multicolumn{2}{c|}{LC-NWC} \\
  \cline{2-5}
  & NTT & INTT & NTT & INTT \\
 \hline
 1024 & 44.5 & 45.5 & 44.5 & 54.5 \\
 \hline
 2048 & 45.0 & 45.8 & 45.0 & 54.2 \\
 \hline
 4096 & 45.5 & 46.2 & 45.5 & 53.8 \\
 \hline
\end{tabular}
\label{table:save_excld1}
\end{table}

\begin{table*}[htbp!]
\centering
\caption{Comparisons between standard LC-INTT and alternative LC-INTT}
\renewcommand{\arraystretch}{1.25}
\begin{tabular}{| c | c | c | c | c | }
 \hline
  & $\#$ multipliers & $\#$ multipliers excl.1 & $\#$ shift operations & $\#$ parameters \\
 \hline
 LC-INTT & $\frac{n}{2}\log n$ & $\frac{n}{2}\log n$ & $n\log n$ & $n-1$ \\
 \hline
 Alt-LC-INTT & $\frac{n}{2}\log n + n$ & $\frac{n}{2}\log n + 1$ & $0$ & $\frac{3n}{2}-1$ \\ 
 \hline
\end{tabular}
\label{table:LC-INTT comp}
\end{table*}

Table \ref{table:LC-INTT comp} shows the comparisons between the standard LC-INTT and the alternative LC-INTT. Although Alt-LC-INTT has $n$ more multipliers compared to LC-INTT, when excluding multiplication by 1, Alt-LC-INTT only has one more multiplier compared to LC-INTT. The main advantage of Alt-LC-INTT is that it doesn't require shift operations for multiplication by $2^{-1}$. Another property worth comparing is the number of parameters that need to be stored in the computation. For LC-INTT, we need to store the 2$n$-th roots from $\psi_{2n}^1$ to  $\psi_{2n}^{n-1}$, that's a total of $n-1$ parameters. For Alt-LC-INTT, we need to store the $n$-th roots from $\omega_n^1$ to  $\omega_n^{(\frac{n}{2}-1)}$ (equivalent to $\psi_{2n}^2$, $\psi_{2n}^4$ to  $\psi_{2n}^{n-2}$), and the $n$ merged multipliers, that's a total of $\frac{3n}{2}-1$ parameters. Therefore, Alt-LC-INTT requires $\frac{n}{2}$ more parameters compared to LC-INTT. The more parameters are used, the more memory allocation is required to compute the result. 

While there are apparent tradeoffs between LC-INTT and Alt-LC-INTT based on the word length of the inputs and the degree of the polynomials, the method of implementation will define which optimized option is better. Since the implementations of these methods aren't usually a one-to-one mapping from algorithm to hardware, different implementations will result in different tradeoffs between the two methods. Hardware implementations in prior works \cite{zhang2020highly,roy2014compact,tan2023parentt} have suggested that LC-NWC with LC-NTT and LC-INTT provides improvements in HE accelerators.


\subsection*{Conclusions}

This lecture note introduced several optimization techniques for NTT-based polynomial modular multiplications. These methods include: zero-padded convolution, negative wrapped convolution, and an improved version of NWC with low-complexity NTT/INTT. 

With low-complexity NTT/INTT, there is no additional polynomial reduction required after the NTT/INTT blocks and no zero-padding is required for both input polynomials. Also, compared to the classical NWC, the pre-processing and post-processing multiplications are eliminated in the low-complexity NWC.

\section*{Acknowledgement}
The authors are grateful to an anonymous reviewer and Prof. Cagatay Candan, the Associate Editor, for their numerous constructive comments.

\section*{Author}
\label{secaut}
\textbf{Sin-Wei Chiu} (chiu0091@umn.edu)
received his bachelor’s degree in electrical engineering from National Central University, Taiwan, in 2020. He is currently pursuing a Ph.D. degree in electrical engineering at the University of Minnesota, Twin Cities. His current research interests include VLSI architecture design, digital signal processing systems, post-quantum cryptography, and homomorphic encryption. 

\textbf{Keshab K. Parhi} (parhi@umn.edu) 
 received his Ph.D. degree in electrical engineering and computer sciences from the University of California, Berkeley, in 1988. He has been with the University of Minnesota, Minneapolis, Minnesota, since 1988, where he is currently the Erwin A. Kelen Chair in Electrical Engineering and a Distinguished McKnight University Professor in the Department of Electrical and Computer Engineering. He has published more than 700 papers, is the inventor of 36 patents, and has authored the textbook VLSI Digital Signal Processing Systems (Wiley, 1999). He served as the editor-in-chief of IEEE Transactions on Circuits and Systems, Part I during 2004 and 2005. He is a Fellow of IEEE, ACM, AIMBE, AAAS, and NAI.

\bibliographystyle{IEEEtran}
\bibliography{refs}



\end{document}